\documentclass[twocolumn,showpacs,preprintnumbers,amsmath,amssymb]{revtex4}

\usepackage{graphicx}
\usepackage{dcolumn}
\usepackage{bm}

\renewcommand{\vec}[1]{\mbox{\boldmath $#1$}}

\begin{document}


\title{Coupled-channels analyses for large-angle quasi-elastic
  scattering in massive systems}
\author{Muhammad Zamrun F.}
\author{K. Hagino}
\affiliation{Department of Physics, Tohoku University,
Sendai 980-8578, Japan}
\author{S. Mitsuoka}
\author{H. Ikezoe}
\affiliation{Advanced Science Research Center, Japan Atomic Energy
  Agency, Tokai, Ibaraki 319-1195, Japan}

\date{\today}

\begin{abstract}
We discuss in detail the coupled-channels approach for the 
large-angle quasi-elastic scattering in massive systems, 
where many degrees of freedom may be involved in the reaction. 
We especially investigate the effects of
single, double and triple phonon excitations on the
quasi-elastic scattering for 
$^{48}$Ti,$^{54}$Cr,$^{56}$Fe,$^{64}$Ni and $^{70}$Zn$+^{208}$Pb
systems, 
for which the experimental cross sections have been measured
recently. 
We show that the present coupled-channels calculations well account 
for the overall width of the experimental barrier distribution for 
these systems. 
In particular, 
it is shown that the calculations taking into 
account single quadrupole phonon excitations in $^{48}$Ti and 
triple octupole phonon excitations in
$^{208}$Pb  reasonably well 
  reproduce the experimental quasi-elastic cross section and 
barrier distribution for the $^{48}$Ti$+^{208}$Pb
  reaction.  
On the other hand, 
$^{54}$Cr,$^{56}$Fe,$^{64}$Ni and $^{70}$Zn$+^{208}$Pb systems seem to
require 
the double quadrupole phonon excitations in the projectiles in order to 
reproduce the experimental data. 
\end{abstract}


\maketitle

\section{Introduction}

It is now well established that the internal structure of colliding
nuclei strongly influences heavy-ion collisions at energies around the
Coulomb barrier. In particular, the coupling to the collective 
excitations (rotation and vibrational states) in the target and projectile
nuclei 
participating in the reaction significantly enhances the fusion
cross sections for intermediate mass systems \cite{bah-tak-98,das-98}. 
Such couplings give rise to a distribution of
the Coulomb barrier \cite{bah-tak-98,das-98,daso-83}, 
which can most easily be visualized for 
reactions involving a deformed nucleus. In this case, the 
nucleus-nucleus potential depends on the orientation angle of the
deformed nucleus with respect to the beam direction. 
Since the orientation angle distributes isotropically 
at the initial stage of the
reaction, so does the potential barrier. 
The concept of barrier distribution can be
extended also to systems with a non-deformed target \cite{daso-83}, 
where the distribution originates from the
coupling between the relative motion and vibrational 
excitations in the colliding nuclei and/or
transfer processes. 
Notice that, although this concept is exact only when the
excitation energy is zero, to a good approximation it holds also 
for systems with a non-zero excitation energy \cite{HTB97,hag-bah-04}. 

In Ref.\cite{row-91}, Rowley {\it et al.} have argued that the barrier
distribution can be directly extracted from a measured fusion cross
section $\sigma_{\rm fus}(E)$, by taking the second derivative of the
product $E\sigma_{\rm fus}(E)$ with respect to the center-of-mass energy
$E$, that is , $D^{\rm fus}=d^2(E\sigma_{\rm fus})/dE^2$. 
This method has stimulated
many high precision measurements of fusion excitation function for
medium-heavy mass systems \cite{das-98,leigh-95}. The
extracted barrier distributions have revealed that the concept indeed holds and
the barrier distribution itself provides a powerful tool for investigating
the effects of channel coupling on heavy-ion fusion reactions at
sub-barrier energies. 
It has also been shown recently that the concept of
barrier distribution is still valid even for relatively heavy systems, such as
$^{100}$Mo$+^{100}$Mo \cite{row-06}.

A similar barrier distribution can also be extracted from quasi-elastic
scattering (a sum of elastic, inelastic and transfer processes) at
backward angles \cite{thim-95,hag-04}, that 
is a good counterpart of the fusion reaction \cite{andres-88}.
In this case, the barrier distribution is  
defined as the first derivative of the ratio of quasi-elastic to the
Rutherford cross sections $d\sigma_{\rm qel}/d\sigma_R$, with respect to
$E$, ${\it i.e.,}$
$D^{\rm qel}=-d(d\sigma_{\rm qel}/d\sigma_R)$/$dE$. 
Since the 
fusion and the quasi-elastic scattering is related to each other 
because of the flux conservation, a similar information can be obtained
from those processes and the similarity between the two
representations for barrier distribution has been shown to hold for several 
intermediate mass systems \cite{thim-95,hag-04,zam-07}. 

Recently, the quasi-elastic barrier distribution has been exploited 
to investigate the entrance channel dynamics for fusion reactions to 
synthesize super-heavy elements \cite{row-061,ntsha-07,ike-06,mitsu-07}. 
It has been shown that the concept of barrier distribution remains
valid even for such very heavy systems once the deep-inelastic 
cross sections are properly taken into account. 
As is expected, the strong channel coupling effects on the barrier 
distribution have been observed. 

In this paper, we carry out a detailed coupled-channels analysis for
large-angle quasi-elastic scattering data for 
$^{48}$Ti,$^{54}$Cr,$^{56}$Fe,$^{64}$Ni and $^{70}$Zn$+^{208}$Pb systems 
leading to super-heavy elements
$Z=104,\,106,\,108,\,110,\,\rm{and}\,112$, respectively
\cite{ike-06,mitsu-07}.  
We especially study the role of 
multi-phonon excitations of the target and projectile nuclei, 
which has been shown to play an important role in 
quasi-elastic scattering for the $^{86}$Kr+$^{208}$Pb system
\cite{ntsha-07}. 

The paper is organized as follows. We briefly explain the coupled-channels 
formalism for quasi-elastic scattering in Sec. II. We present the
results of our  
systematic analysis in Sec. III.  
We then summarize the paper in Sec. IV. 
 
\section{Coupled-channels Formalism for Large Angle Quasi-elastic Scattering}

In this section, we briefly describe the coupled-channels formalism
for large angle quasi-elastic scattering which includes the effects of the
vibrational excitations of the colliding nuclei.
The total Hamiltonian of the system is assumed to be
\begin{eqnarray}
H&=&-\frac{\hbar^2}{2\mu}\nabla^2+V^{(0)}_N(r)+\frac{Z_PZ_Te^2}{r} \nonumber \\
&&+H_{\rm exct}
+V_{\rm coup}(\vec{r},\xi_P,\xi_T),
\label{eq-multi-1}
\end{eqnarray}
where $\vec{r}$ is the coordinate of the relative motion between the 
target and the projectile nuclei, $\mu$ is the reduced mass and
$\xi_T$ and $\xi_P$ represent the coordinate of the vibration 
in the target and the projectile nuclei, respectively. 
$Z_P$ and $Z_T$ are the atomic number of
the projectile and the target, respectively, and 
$V^{(0)}_N$ is the bare 
nuclear potential, which we assume to have a Woods-Saxon shape. It 
consists of the real and imaginary parts,  
$V^{(0)}_N(r)=V_0(r)+iW_0(r)$. 
$H_{\rm exct}$ describes the
excitation spectra of the target and projectile nuclei, while 
$V_{\rm coup}(\vec{r},\xi_P,\xi_T)$ is the potential for the 
coupling between the 
relative motion and the
vibrational motions of the target and projectile nuclei. 

In the iso-centrifugal approximation
\cite{bah-tak-98,hag-99,es-lan-pri-87}, where 
the the angular momentum of the relative motion in each
channel is replaced with the total angular momentum $J$ 
(in the literature, this approximation is also 
referred to as the rotating frame approximation or the no-Coriolis
approximation), the coupled-channels
equations derived from the Hamiltonian (\ref{eq-multi-1}) read 
\begin{eqnarray}
&&\bigg[-\frac{\hbar^2}{2\mu}\frac{d^2}{dr^2}+\frac{J(J+1)\hbar^2}{2\mu
  r^2}+V^{(0)}_N(r)+\frac{Z_PZ_Te^2}{r}\nonumber\\
&&-E+\epsilon_n\bigg]u_n(r)+\sum_{n'} V_{nn'}(r)u_{n'}(r)=0
\label{eq-multi-11}
\end{eqnarray}
where $\epsilon_n$ is the eigen-value of the operator $H_{\rm exct}$
for the $n$-th channel. $V_{nn'}(r)$ is 
the matrix elements for the coupling potential $V_{\rm coup}$. 

In the calculations presented below, we use the method 
of the computer code {\tt CCFULL} \cite{hag-99} and replace the 
vibrational coordinates $\xi_P$ and $\xi_T$ in the coupling potential 
$V_{\rm coup}$ with the dynamical excitation operators
$\hat{O}_{P}$ and $\hat{O}_{T}$. The coupling potential is then
represented as 
\begin{equation}
V_{\rm coup}(r,\hat{O}_P,\hat{O}_T)
=V_C(r,\hat{O}_P,\hat{O}_T)+V_N(r,\hat{O}_P,\hat{O}_T),
\label{eq-multi-2}
\end{equation}
\begin{eqnarray}
V_C(r,\hat{O}_P,\hat{O}_T)&=&\left(
\frac{3R_P^{\lambda_P}\hat{O}_{P}}{(2\lambda_P+1)r^{\lambda_P}}
+\frac{3R_T^{\lambda_T}\hat{O}_T}{(2\lambda_T+1)r^{\lambda_T}}\right)
\nonumber\\
&&\times \frac{Z_PZ_Te^2}{r},\label{eq-multi-3}\\
V_{N}(r,\hat{O}_P,\hat{O}_T)&=&\frac{-V_0}{\left[1+\textrm{exp}\left(\frac{
[r-R_0-(R_P\hat{O}_P+R_T\hat{O}_T)]}
{a}\right)\right]} \nonumber \\
&&-V_N^{(0)}(r). 
\label{eq-multi-4}
\end{eqnarray}
Here, $\lambda_P$ and $\lambda_T$ denote the multipolarity of the vibrations
in the projectile and the target nuclei, respectively. 
We have subtracted $V_N^{(0)}(r)$ in Eq.~(\ref{eq-multi-4}) in order
to avoid the double counting. 

If we truncate the phonon space 
up to the triple phonon states (that is, $n$=0,1,2, and
3), the matrix elements of 
the excitation operator $\hat{O}$ in Eqs.~(\ref{eq-multi-3}) 
and (\ref{eq-multi-4}) are given by 
\begin{eqnarray}
O_{nn'}=\frac{1}{\sqrt{4\pi}}
\left[\begin{array}{cccc}
0&\beta&0&0\\
\beta&0&\sqrt{2}\beta&0\\
0&\sqrt{2}\beta&0&\sqrt{3}\beta \\
0&0&\sqrt{3}\beta&0
\end{array}\right]
\label{eq-multi-5}
\end{eqnarray}
where $\beta$ is the deformation parameter, that can be 
estimated from a measured electric transition probability from the
single phonon state ($n$=1) to the ground state ($n$=0). 
We have assumed the harmonic oscillator model for the vibrations,
where $\epsilon_n$ in Eq.~(\ref{eq-multi-11}) is given by 
$\epsilon_n = n\,\hbar\omega$. 

The coupled-channels equations, Eq.~(\ref{eq-multi-11}), are solved with
the scattering boundary condition for $u_n(r)$, 
\begin{eqnarray}
u_n(r)&\rightarrow&
\frac{i}{2}\left\{H_J^{(-)}(k_nr)\delta_{n,n_i}-\sqrt{\frac{k_i}{k_n}}
S_n^JH_J^{(+)}(k_nr)\right\}, \nonumber \\
&&~~~~~~~(r\to\infty)
\label{eq-scattwave}
\end{eqnarray} 
where $S_n^J$ is the nuclear $S$ matrix. $H_J^{(-)}(kr)$ and
$H_J^{(-)}(kr)$ are the incoming and the outgoing Coulomb wave
functions, respectively. The channel wave number $k_n$ is given by
$\sqrt{2\mu \,(E-\epsilon_n)/\hbar^2}$, and $k_i=k_{n_i}=\sqrt{2\mu
  E/\hbar^2}$. 
The scattering angular distribution for the channel $n$
is then given by \cite{es-lan-pri-87}
\begin{equation}
\frac{d\sigma_n}{d\Omega}=\frac{k_n}{k_i}\lvert f_n(\theta)\rvert^2
\end{equation}
with 
\begin{eqnarray}
f_n(\theta)=\sum_Je^{[\sigma_J(E)+\sigma_J(E-\epsilon_n)]}
\sqrt{\frac{2J+1}{4\pi}}Y_{J0}(\theta)\nonumber\\
\times \frac{-2i\pi}{\sqrt{k_ik_n}}(S_n^J-\delta_{n,n_i})
  +f_C(\theta)\delta_{n,n_i}
\end{eqnarray}
where $\sigma_J(E)$ and $f_C(\theta)$ are the the Coulomb phase shift
and the Coulomb scattering amplitude, respectively. The differential
quasi-elastic cross section is then calculated to be 
\begin{equation}
\frac{d\sigma^{\rm qel}}{d\Omega}=\sum_n\frac{d\sigma_n}{d\Omega}
\end{equation}
We will apply this 
formalism in the next section to analyze the  
quasi-elastic scattering data of 
$^{48}$Ti,$^{54}$Cr,$^{56}$Fe,$^{64}$Ni, and $^{70}$Zn$+^{208}$Pb
systems.  

\section{Comparison with experimental data : effects of 
multi-phonon excitations}

In this section, we present the results of our detailed coupled-channels
analysis for quasi-elastic scattering data of 
$^{48}$Ti,$^{54}$Cr,$^{56}$Fe,$^{64}$Ni, and $^{70}$Zn$+^{208}$Pb systems
\cite{ike-06,mitsu-07}.
The  calculations are performed with a version \cite{hag2} of the
coupled-channels code {\tt CCFULL} \cite{hag-99}.
Notice that the iso-centrifugal approximation employed in this code 
works well for
quasi-elastic scattering at backward angles \cite{hag-04}. In the code,
the regular boundary condition is imposed at the origin instead of the
incoming wave boundary condition. 
We discuss the stability of the numerical calculations 
for the massive systems in Appendix A. 

The surface diffuseness of the real part of the nuclear
potential is taken to be $a=0.63$ fm, as suggested by recent studies 
on deep sub-barrier quasi-elastic and Mott scattering 
\cite{washi-06,gasques-07,hinde-07}, while the radius parameter to be
$r_0=1.22$ fm for all the systems. 
Notice that a similar value for $a$ has been used also in the 
analysis of the recent experimental data for quasi-elastic 
scattering in the $^{86}$Kr + $^{208}$Pb system \cite{ntsha-07}. 
The depth parameter, $V_0$, is adjusted 
in order to reproduce the experimental quasi-elastic cross sections for each
system. 
The optimum values of the depth parameter and the
resultant Coulomb barrier height are summarized in Table~\ref{pot-par}. As
usually done, we use a short range imaginary potential with 
$W_{0}=30$ MeV, $r_{w}=1.0$ fm and $a_w=0.3$ fm to simulate the
compound nucleus formation.  
The results are insensitive to these parameters 
as long as the imaginary part of the potential is well 
confined inside the Coulomb barrier. 
The excitation
energy and the corresponding deformation parameter for the single 
phonon excitation in each nucleus 
included in the calculations are given in Table~\ref{ene-def}. 
The latter quantity is taken from Refs. \cite{spear-02,raman-01}.  
The radius of the target and
the projectile are taken to be $R_T=1.2A_T^{1/3}$ and  $R_P=1.2A_P^{1/3}$,
respectively, in order to be consistent with the deformation
parameters \cite{spear-02,raman-01}.  
All the calculations shown below 
are performed at the scattering angle of 
$\theta_{\rm c.m.}=170^\circ$.  
We plot the quasi-elastic cross sections and barrier distributions 
as a function of 
the effective energy defined by \cite{thim-95,hag-04}
\begin{equation}
E_{\rm eff}=2E\frac{\rm sin(\theta/2)}{1+\rm sin(\theta/2)}, 
\end{equation}
which takes into account the centrifugal energy. 
We calculate the quasi-elastic barrier
distributions from the cross sections 
in a similar way as the one used to obtain 
the experimental barrier distributions \cite{mitsu-07}. 
Namely, we use 
the point difference formula with the energy step of $\Delta E$=0.25
MeV and then smooth the resultant barrier distribution with the 
Gaussian function with the full width at half maximum (FWHM) of 1.5 
MeV. We have checked that the shape of the barrier distribution does
not change significantly even if we use a larger energy step for the 
point difference formula, e.g. $\Delta E$=0.5 MeV. 

\begin{table}
\caption{\label{pot-par}The 
depth parameter for the real part of the nuclear potential 
for the $^{48}$Ti,$^{54}$Cr,$^{56}$Fe,$^{64}$Ni, and $^{70}$Zn$+^{208}$Pb
systems. The radius and the diffuseness parameters are taken to be 
$r_0$=1.22 fm and $a$=0.63 fm, respectively, for all the systems. 
The resultant barrier height energy $V_B$ is also listed.}
\begin{ruledtabular}
\begin{tabular}{ccc}
System&$V_0$(MeV)&$V_B$(MeV)\\
\hline
$^{48}$Ti+$^{208}$Pb & 88.90 & 190.50\\
$^{54}$Cr+$^{208}$Pb & 91.70 & 205.50\\
$^{56}$Fe+$^{208}$Pb & 92.85 & 222.50\\
$^{64}$Ni+$^{208}$Pb & 95.10 & 236.25\\
$^{70}$Zn+$^{208}$Pb & 108.2 & 249.30\\
\end{tabular}
\end{ruledtabular}
\end{table}
\begin{table}
\caption{\label{ene-def}The properties of the single phonon states 
included in the present coupled-channels calculations. 
$\hbar\omega$ and $\beta$ are the excitation energy and the dynamical 
deformation parameter, respectively. 
} 
\begin{ruledtabular}
\begin{tabular}{cccc}
Nucleus&$I^\pi$&$\hbar\omega$ (MeV)&
$\beta$\\
\hline
$^{208}$Pb&$3^-$ & $2.614$ & $0.110$ \footnotemark[1]\\
$^{48}$Ti &$2^+$ & $0.983$ & $0.269$ \footnotemark[2]\\
$^{54}$Cr &$2^+$ & $0.834$ & $0.250$ \footnotemark[2]\\
$^{56}$Fe &$2^+$ & $0.846$ & $0.239$ \footnotemark[2]\\
$^{64}$Ni &$2^+$ & $1.346$ & $0.179$ \footnotemark[2]\\
$^{70}$Zn &$2^+$ & $0.884$ & $0.228$ \footnotemark[2]\\
\end{tabular}
\end{ruledtabular}
\footnotetext[1]{taken from Ref.~\cite{spear-02}.}
\footnotetext[2]{taken from Ref.~\cite{raman-01}.}
\end{table}

\begin{figure}
\includegraphics[width=.45\textwidth]{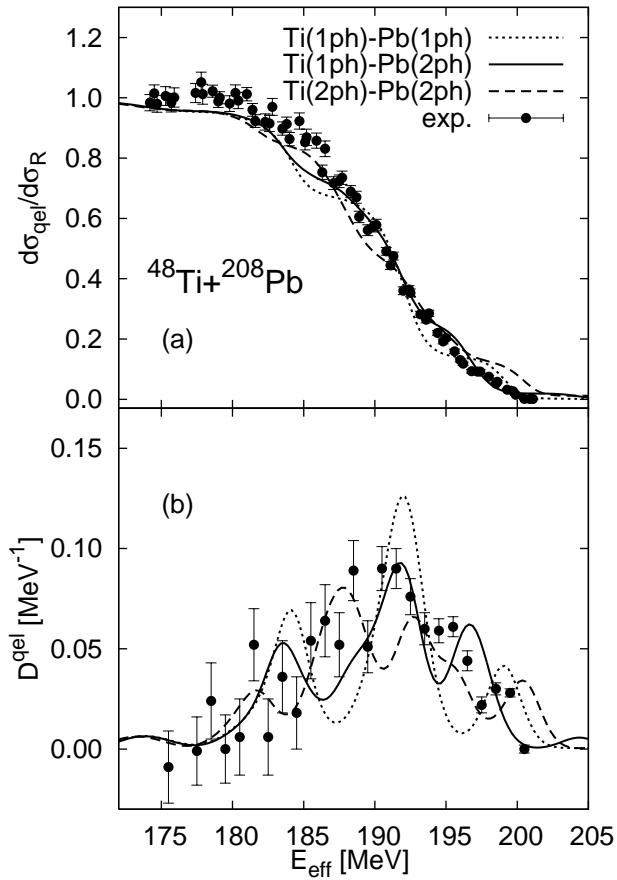}
\caption{Effects of multi-phonon excitations on the quasi-elastic scattering 
cross section (upper panel) and on the quasi-elastic barrier distribution
(lower panel) for the $^{48}$Ti+$^{208}$Pb system. 
The dotted line is the result of the coupled-channels calculations
including coupling to the one quadrupole phonon state in the
projectile 
and the one octupole phonon states in the target nucleus, while the
solid line is obtained by including the coupling in addition to 
the two octupole phonon state in the
target nucleus. 
The dashed line is the result of double quadrupole phonon couplings in
the projectile and the double octupole phonon couplings in the target
nucleus. 
The experimental data are taken from  
Ref. \cite{mitsu-07}.}
\label{fig-tipb-1}
\end{figure}

\subsection{Effect of double phonon excitations}

\begin{figure*}
\includegraphics[width=.79\textwidth]{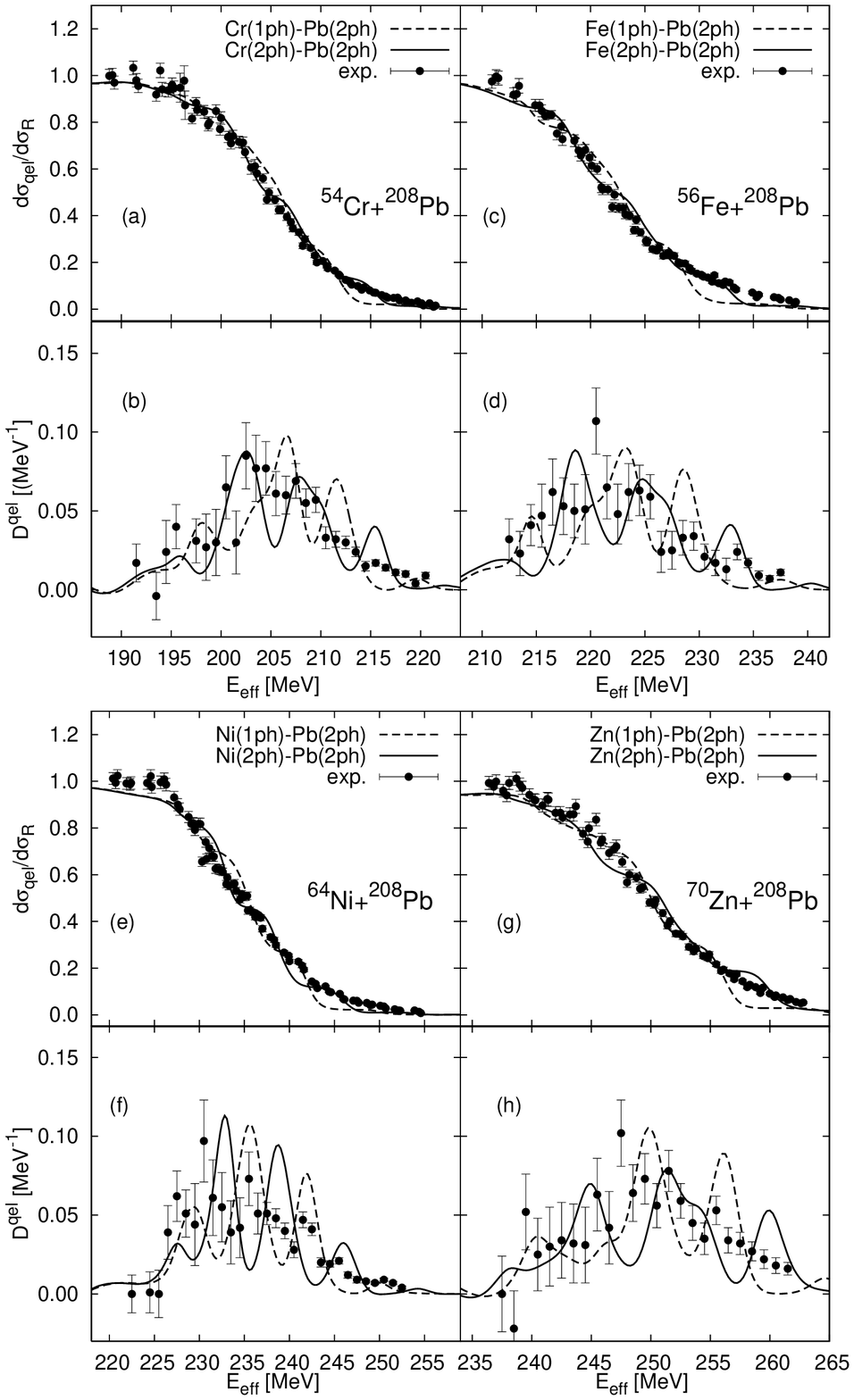}
\caption{
The quasi-elastic scattering 
cross sections ((a), (c), (e), and (g)) and the quasi-elastic barrier
distributions ((b), (d), (f), and (h)) for the
$^{54}$Cr,$^{56}$Fe,$^{64}$Ni and $^{70}$Zn$+^{208}$Pb systems
obtained with two coupling schemes as indicated in the insets. 
The dashed line is obtained by including the one quadrupole phonon
state in the projectile nuclei while the solid line is obtained with
the double phonon couplings. The double octupole phonon excitations in
the target nucleus is included in all the calculations. 
The experimental data are taken from Ref. \cite{mitsu-07}.}
\label{fig-crfeniznpb-1}
\end{figure*}

Let us first discuss the effect of double octupole phonon excitations
in the $^{208}$Pb target. 
Such excitations 
have been shown to play a significant role 
in the sub-barrier fusion reaction between $^{16}$O and $^{208}$Pb 
nuclei \cite{das-97,morton-99}. 

\begin{figure}
\includegraphics[width=.45\textwidth]{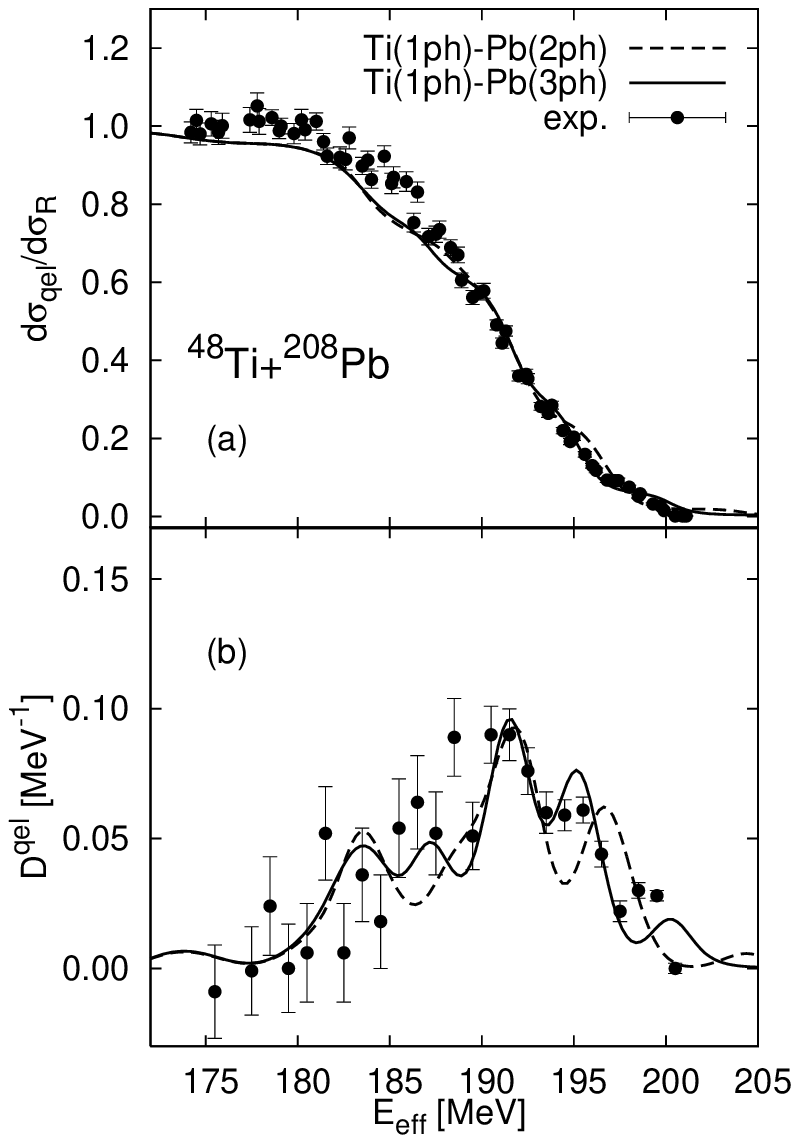}
\caption{Effects of triple phonon excitations on the quasi-elastic scattering 
cross section (upper panel) and on the quasi-elastic barrier
distribution (lower panel) for the $^{48}$Ti+$^{208}$Pb system. 
The dashed line is the result of the coupled-channels calculations 
taking into account the coupling to the one phonon state in the projectile and
the two phonon states in the target nuclei. The solid line
is obtained by including the coupling to the one phonon state in the
projectile and the three phonon states in 
the target. The experimental
data are taken from Ref. \cite{mitsu-07}.} 
\label{fig-tipb-2}
\end{figure}

The dotted line in Fig.~\ref{fig-tipb-1} shows the 
result of the coupled-channels calculation 
for the $^{48}$Ti + $^{208}$Pb system 
obtained by taking into 
account the 
coupling to the single octupole phonon state 
in the target nucleus, $^{208}$Pb, and the single quadrupole phonon 
state in the projectile nucleus, $^{48}$Ti. 
The mutual excitations in the projectile and the target nuclei 
are fully taken into account in this calculation as well as in all the
other calculations presented in this paper. 
Figs.~\ref{fig-tipb-1}(a) and 
\ref{fig-tipb-1}(b) show the ratio of the quasi-elastic to the
Rutherford cross sections, $d\sigma_{\rm qel}/d\sigma_R$, and the
quasi-elastic barrier distribution, $D^{\rm qel}$, respectively.
Although the overall width of the barrier distribution is reproduced 
reasonably well with this calculation, the detailed structure is 
somewhat inconsistent with the experimental data. 
The situation is similar even when we include the double quadrupole
phonon state in the projectile while keeping the single octupole
phonon coupling in the target nucleus (not shown). 
We then investigate the effect of the double octupole phonon couplings
in the target nucleus. The solid and the dashed lines in 
Fig.~\ref{fig-tipb-1}  show the results with 
the single and the double quadrupole phonon
excitations in the projectile, respectively, where as the double 
octupole phonon coupling in the target is included in both the
calculations. The former calculation reproduces both the cross
sections and the barrier distribution reasonably well, although the
latter calculation somehow worsens the agreement. 
This clearly suggests that the double
octupole phonon excitations in the target nucleus is 
important in the quasi-elastic $^{48}$Ti+$^{208}$Pb scattering. 
We summarize the $\chi^2$ value of our calculations in Table~\ref{chi-1}. 

Since the coupling to the one quadrupole phonon state in the projectile
and the two
octupole phonon states in the target reasonably well reproduce the
experimental quasi-elastic scattering data for the $^{48}$Ti+$^{208}$Pb system,
one may expect that the same coupling scheme accounts for the
experimental 
data for the other systems, 
$^{54}$Cr,$^{56}$Fe,$^{64}$Ni and $^{70}$Zn$+^{208}$Pb. 
The results of the coupled-channels calculations with this
coupling scheme is shown by the dashed line in
Fig.~\ref{fig-crfeniznpb-1}. 
Figs.~\ref{fig-crfeniznpb-1}(a), (c), (e), and (g) 
are for 
the quasi-elastic cross sections 
for the $^{54}$Cr,$^{56}$Fe,$^{64}$Ni and $^{70}$Zn$+^{208}$Pb
systems, respectively, 
while Figs.~\ref{fig-crfeniznpb-1}(b), (d), (f), and (h) are 
for the quasi-elastic barrier distributions.
One can clearly see that these 
calculations underestimate the experimental cross sections 
at high energies, although the experimental barrier distributions
themselves are reproduced reasonably well. 
We repeat the same
calculations by including the coupling up to the double quadrupole phonon
states in the projectile, in addition to the double octupole phonon states
in the target nucleus. These results are shown by the solid in
Fig.~\ref{fig-crfeniznpb-1}. 
The agreement with the experimental data is considerably improved,
especially for the quasi-elastic cross sections. See Table~\ref{chi-2} for the 
$\chi^2$ values.  
It is thus evident that 
the coupling to the double
quadrupole phonon states in the projectile 
is needed in
order to explain the experimental data for the 
 $^{54}$Cr,$^{56}$Fe,$^{64}$Ni and $^{70}$Zn$+^{208}$Pb reactions. 

The reason why the double quadrupole phonon coupling is not necessary
for the $^{48}$Ti projectile while it is for the heavier projectiles 
is not clear at this moment. 
This might reflect some ambiguity of the Monte Carlo
reaction simulation code {\tt LINDA} \cite{LINDA} which was 
used to subtract the
deep-inelastic component from the experimental yields at backward
angles \cite{mitsu-07}. 
Clearly, a further investigation is still necessary concerning the 
effect of deep inelastic scattering on quasi-elastic scattering in
massive systems \cite{row-061,ntsha-07,ike-06,mitsu-07}.

\begin{figure*}
\includegraphics[width=.79\textwidth]{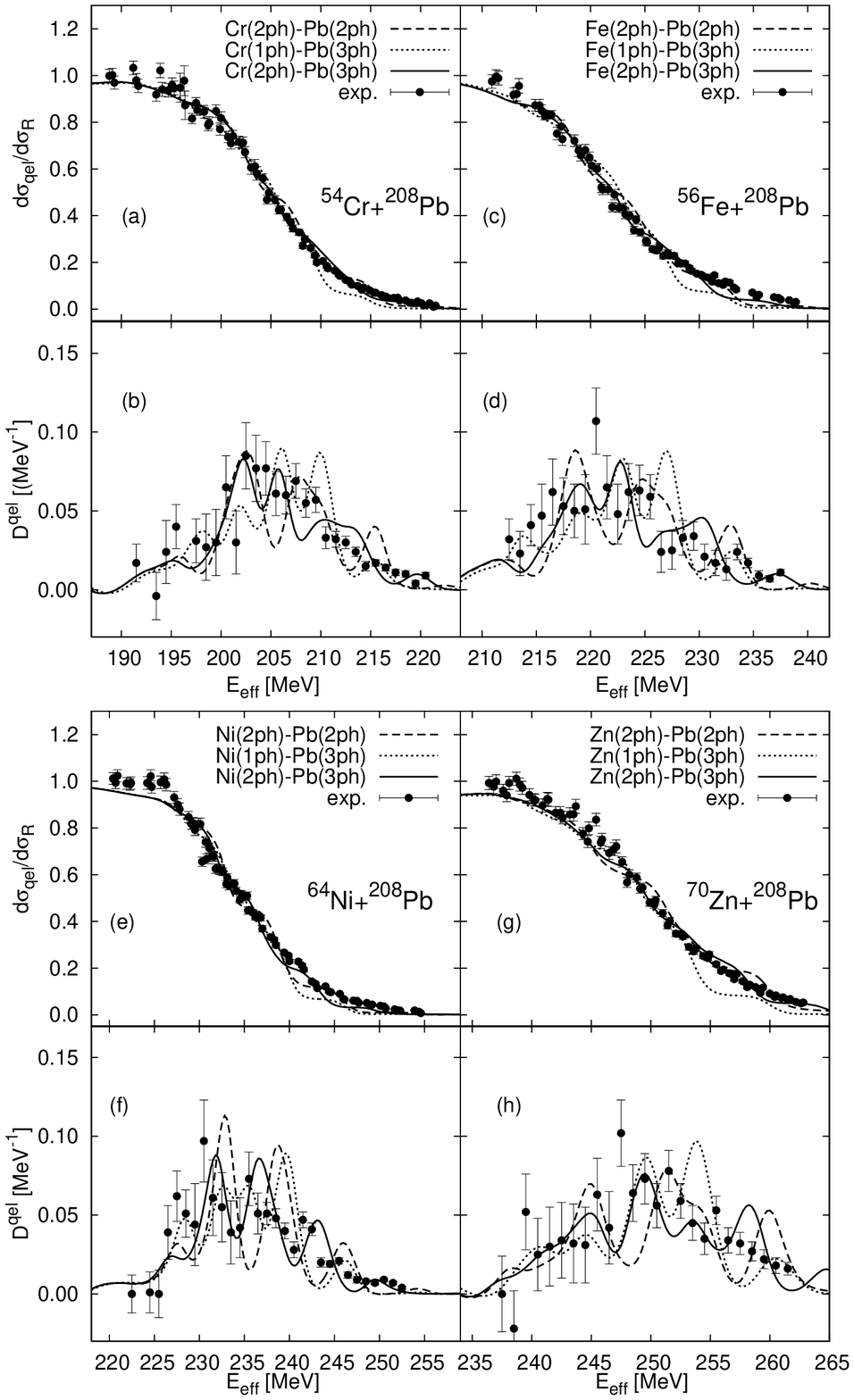}
\caption{Effects of triple phonon excitations on the quasi-elastic 
cross sections ((a), (c), (e), and (g)) and on the quasi-elastic barrier
distributions ((b), (d), (f), and (h)) for the
$^{54}$Cr,$^{56}$Fe,$^{64}$Ni and $^{70}$Zn$+^{208}$Pb systems.  
The dashed line is the same as the solid line in Fig.~\ref{fig-crfeniznpb-1} 
while the dotted line is the results of the 
calculations taking the coupling to the triple octupole phonon in the target 
and the one quadrupole phonon state in the projectile nucleus into
account. The solid line is obtained by including the coupling to the double 
quadrupole phonon states 
in the projectile and the triple otcupole phonon states in the target
nucleus. The experimental data are taken from Ref. \cite{mitsu-07}.}
\label{fig-crfeniznpb-2}
\end{figure*}

\begin{table}
\caption{\label{chi-1}
The value of $\chi^2$ for the quasi-elastic cross sections for the 
$^{48}$Ti$+^{208}$Pb system obtained with the coupled-channels
calculations with various coupling schemes. The coupling schemes are 
denoted as $[n_2,n_3]$, where $n_2$ is the number of quadrupole phonon 
excitation in the projectile nucleus while $n_3$ the number of 
octupole phonon in the target nucleus. 
}
\begin{ruledtabular}
\begin{tabular}{ccccc}
System&[1,1]&[1,2]&[1,3]&[2,2]\\
\hline
$^{48}$Ti+$^{208}$Pb & 19.15 & 9.82 & 7.12 & 37.12\\
\end{tabular}
\end{ruledtabular}
\end{table}

\begin{table}
\caption{\label{chi-2}
Same as Table \ref{chi-1}, but for the $^{54}$Cr,$^{56}$Fe,$^{64}$Ni 
and $^{70}$Zn$+^{208}$Pb systems.}
\begin{ruledtabular}
\begin{tabular}{ccccc}
System&[1,2]&[1,3]&[2,2]&[2,3]\\
\hline
$^{54}$Cr+$^{208}$Pb & 52.47 & 49.80 & 20.61 & 11.78\\
$^{56}$Fe+$^{208}$Pb & 28.46 & 28.36 & 10.44 & 10.28\\
$^{64}$Ni+$^{208}$Pb & 57.45 & 61.43 & 32.21 & 30.64\\
$^{70}$Zn+$^{208}$Pb & 26.52 & 24.81 & 11.36 & 6.87\\
\end{tabular}
\end{ruledtabular}
\end{table}

\subsection{Effect of triple phonon excitations}

In the previous subsection, we have shown that the double octupole phonon
excitations in the $^{208}$Pb target play an important role 
in quasi-elastic scattering for the systems considered in this paper. 
However, the calculated quasi-elastic barrier distributions have a much more
prominent peak than the experimental distribution at high energies. 
Since it has been shown in Refs. \cite{row-061,ntsha-07} that 
the triple octupole phonon excitations of the $^{208}$Pb
play a significant role in 
the large-angle quasi-elastic
scattering between $^{86}$Kr and $^{208}$Pb nuclei, it is 
intriguing to investigate such effects in the present systems as
well. 

The results of the coupled-channels calculations including the
coupling to the triple octupole phonon states in $^{208}$Pb 
for the $^{48}$Ti+$^{208}$Pb reaction is presented in Fig.~\ref{fig-tipb-2}. 
The dashed line is the same as the solid line in
Fig.~\ref{fig-tipb-1}, that is the result of single phonon in
$^{48}$Ti and double phonon in $^{208}$Pb. 
The solid line denotes the results of the triple phonon 
coupling in the target in addition to the single phonon 
in the projectile. 
By including the triple octupole phonons in the target nucleus, 
the quasi-elastic cross sections are improved slightly (see also
Table~\ref{chi-1}).  
On the other hand, one can see that the agreement 
for the barrier distribution with the experimental data 
is much more improved by the triple phonon coupling. 

\begin{figure}
\includegraphics[width=.45\textwidth]{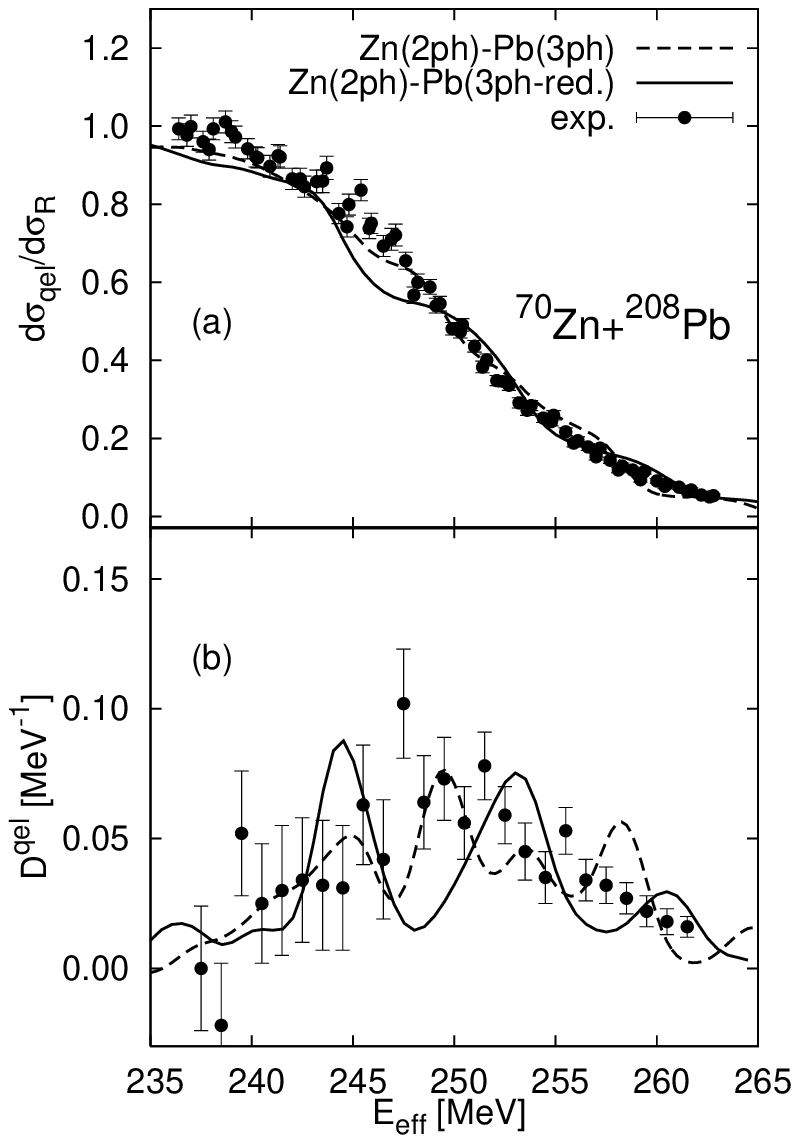}
\caption{Effect of anharmonic octupole phonon excitations in
  $^{208}$Pb on 
(a) the quasi-elastic cross section and 
(b) the quasi-elastic barrier distribution 
for the $^{70}$Zn$+^{208}$Pb reaction. 
The solid line is the coupled-channels calculation 
obtained by reducing the coupling strengths to multi-phonon states,
while the dashed line denotes the results in the 
  harmonic limit. The experimental data are taken from Ref. \cite{mitsu-07}.} 
\label{fig-ahv}
\end{figure}

The results for the other systems, 
$^{54}$Cr,$^{56}$Fe,$^{64}$Ni and $^{70}$Zn$+^{208}$Pb reactions, 
are shown in Fig.~\ref{fig-crfeniznpb-2}. Figs. \ref{fig-crfeniznpb-2}(a), 
(c), (e) and (g) are 
for the quasi-elastic cross sections, while Figs.~\ref{fig-crfeniznpb-2}(b), 
(d), (f) and (h) for the quasi-elastic barrier distributions. 
Let us first discuss the calculations with the single phonon
excitation in the projectile. The  dotted line in the figures 
is obtained by taking
the coupling to the single phonon state in the projectile and
the triple octupole phonon excitations in the target. 
This calculation 
underestimates the quasi-elastic cross sections at high
energies and the obtained barrier distribution is inconsistent with the
experimental data. Therefore, the previous results shown in Fig. 
~\ref{fig-crfeniznpb-1} is not improved even if the triple phonon 
excitations in the target is taken into account as long as only the
single phonon excitation is considered for the projectile nucleus. 
The results with 
the double phonon couplings in the projectile 
together with the triple phonon excitations in the target  
are then shown by the solid
line in the figure. 
For comparison, we also show by the dashed line 
the results of the double phonon excitations in both the projectile
and the target nuclei, which is the same as the solid line in
Fig. ~\ref{fig-crfeniznpb-1}. 
One can observe that the inclusion of the triple octupole phonon
excitations in the $^{208}$Pb 
somewhat improves the agreement 
between the calculations and the experimental data for both 
the quasi-elastic cross sections and the barrier distributions (see
also Table~\ref{chi-2}). 

In Ref.\cite{ntsha-07}, Ntshangase {\it et al.} 
reduced the coupling strength of $(3^-)\rightarrow(3^-)^2$ 
states in $^{208}$Pb by a factor of $(0.6)$ and 
that of $(3^-)^2\rightarrow(3^-)^3$ by $(0.6)^2$ 
in order to explain the  
experimental barrier distribution for the $^{86}$Kr$+^{208}$Pb reaction. 
In order to 
see whether such reduction of the coupling strengths improves 
the agreement between the coupled-channels calculations and the
experimental data for the present systems, 
we repeat the calculations 
by including those effects for the $^{70}$Zn$+^{208}$Pb system. 
The results are shown in Fig.~\ref{fig-ahv}. 
The solid is 
obtained by reducing the coupling strengths as 
Ntshangase {\it et al.} did, 
while the dashed line is the same as the solid line in
Figs.~\ref{fig-crfeniznpb-2}(g) and (h), that is 
obtained by assuming the harmonic limit. 
In both cases, we take into
account the coupling to the double quadrupole phonon excitations in the
projectile nucleus. 
One can see that the harmonic model 
leads to a better agreement with the experimental data 
both for the cross sections and the barrier distribution, as compared 
to the anharmonic calculation. 
The difference between Ref. \cite{ntsha-07} and the present
calculation concerning the role of anharmonicity may originate from
the fact that  Ref. \cite{ntsha-07} used a smaller value for 
$R_T$ (=1.06$A_T^{1/3}$ fm) and thus a larger value for $\beta_3$
(=0.16).  In order to clarify the role of anharmonicity of
multi-phonon excitations in quasi-elastic scattering in massive
systems, 
it would be required to take into account also the 
reorientation terms \cite{hag-97,hag-971,zam-07}. It is beyond the 
scope of this paper, and we will leave it for a future study. 

\begin{figure}
\includegraphics[width=.45\textwidth]{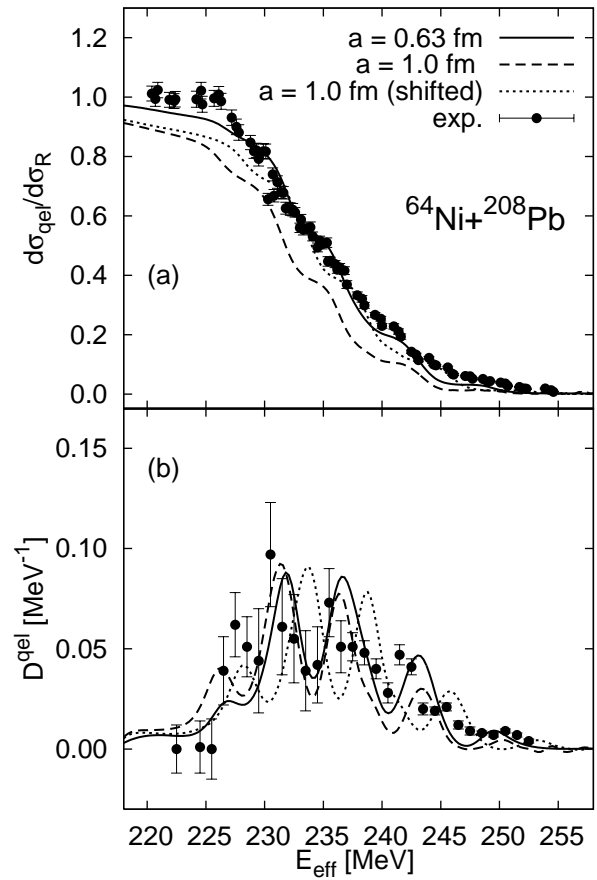}
\caption{Comparison of the experimental data with the coupled-channels 
calculations obtained using different values of the surface diffuseness of 
the nuclear potential for $^{64}$Ni$+^{208}$Pb reaction for 
(a) the quasi-elastic cross section and 
(b) the quasi-elastic barrier distribution. 
The solid and the dashed lines is obtained using the surface 
diffuseness of the nuclear potential $a=0.63$ fm and $a=1.0$ fm,
respectively. The dotted line is the results obtained by shifting the barrier 
height by around $+\, 2.40$ MeV for the calculation using $a=1.0$ fm.  
Experimental data is taken from Ref. \cite{mitsu-07}.} 
\label{fig-diff}
\end{figure}

\subsection{Surface diffuseness of the nuclear potential}

We next discuss the dependence of the quasi-elastic scattering 
on the surface diffuseness parameter of the nuclear potential. 
The standard value for the diffuseness parameter is around 0.63 fm 
 \cite{brogwin-91,criswin-76,satc-79}. 
Recently, systematic studies on 
quasi-elastic scattering as well as Mott scattering at deep sub-barrier 
energies have revealed that the surface region of the nuclear 
potential is indeed consistent with the standard value of the 
surface diffuseness parameter ~\cite{washi-06,gasques-07,hinde-07}. 
On the other hand, 
it has been known for some time 
that the recent high precision data of sub-barrier fusion cross 
sections require a larger value of surface diffuseness 
parameter, ranging 
between 0.75 and 1.5 fm \cite{NBD04}. 
Since the large-angle quasi-elastic scattering around the Coulomb
barrier may probe both the surface region and the inner part of the 
nuclear potential, it is interesting to study the sensitivity of 
quasi-elastic cross sections and barrier distributions 
to the surface diffuseness parameter. 

For this purpose, as an example, we repeat 
the coupled-channels calculation for the 
the $^{64}$Ni$+^{208}$Pb reaction using the nuclear potential with 
$a$=1.0 fm. 
We readjust the depth and the radius 
parameters to be $V_0=160.70$ MeV and $r_0=1.10$ fm, respectively, 
so that the barrier height remains the same as the one listed in Table
I. We include 
the coupling to the double phonon states in the projectile and 
the triple phonon states in the target. 
Fig.~\ref{fig-diff} compares the results with $a$=0.63 fm (the solid
line) to the one with $a$=1.0 fm (the dashed line). 
One sees that the calculations with $a$=1.0 fm 
underestimate the quasi-elastic cross section, although 
the shape of barrier distribution itself is similar to the one obtained
with $a=0.63$ fm. 
The dotted line is obtained with the same value of surface 
diffuseness parameter $a$=1.0 fm as the one for the dashed line, 
but by changing 
the depth parameter $V_0$ so that the resultant barrier height is 
higher by 2.4 MeV. 
This calculation now reproduces the experimental quasi-elastic cross 
sections at energies larger than $E_{\rm c.m.}=230$ MeV 
reasonably well, but below this energy 
the cross sections are underestimated. 
Therefore, it seems difficult to reproduce the experimental
quasi-elastic cross sections with the diffuseness parameter of 
$a$=1.0 fm at energies below and
above the Coulomb barrier simultaneously.  
We have checked that the situation is similar for the other systems, 
$^{48}$Ti,$^{54}$Cr,$^{56}$Fe and $^{70}$Zn$+^{208}$Pb. 
This result clearly indicates that the standard value of surface 
diffuseness, $a=0.63$ fm, 
is preferred by 
the experimental quasi-elastic data 
for the systems studied in this paper.

\section{Summary}

We have performed a detailed coupled-channels analysis for
large-angle quasi-elastic scattering of the
$^{48}$Ti,$^{54}$Cr,$^{56}$Fe,$^{64}$Ni and $^{70}$Zn$+^{208}$Pb
systems, where their experimental 
barrier distributions have been extracted 
recently. Our coupled-channels calculations with multi-phonon 
excitations in the colliding nuclei reproduce the experimental 
quasi-elastic cross sections as well as the barrier distributions, 
indicating clearly that 
the coupled-channels approach still works even for massive 
systems \cite{row-06}. 
It was crucial to subtract properly the 
deep-inelastic components from the total backward-angle cross sections 
in order to reach these agreements between the calculations and the 
experimental data. 

In more details, 
the calculation with the single
quadrupole phonon excitation in 
$^{48}$Ti and the triple 
octupole phonon excitations in $^{208}$Pb 
reproduces reasonably well the experimental data 
for the $^{48}$Ti$+^{208}$Pb system. 
On the other hand, for the 
$^{54}$Cr,$^{56}$Fe,$^{64}$Ni and $^{70}$Zn$+^{208}$Pb systems, 
we found that the
coupling to the double quadrupole phonon excisions in the projectile
nucleus in addition to the coupling to the triple octupole phonon in the
target nucleus seems to be needed to fit the experimental data. 
These results suggest that the triple octupole  phonon 
excitations in the $^{208}$Pb nucleus plays an
important role in describing the experimental data for the quasi-elastic
cross section and the quasi-elastic barrier distribution for 
the present massive systems. This is consistent with the previous 
finding for the $^{86}$Kr + $^{208}$Pb system \cite{ntsha-07}. 

Although our calculations well reproduce the gross features 
of the experimental barrier distributions, 
higher precision data are still required in order to study the 
detailed structure of the barrier distributions, 
especially the role of multi octupole phonon 
states in $^{208}$Pb. 
From the theoretical side, a further detailed investigation 
will also be necessary, 
taking into account the anharmonicity of the multi-phonon excitations.  

\begin{acknowledgments}
This work was partly supported by The 21st Century Center of
Excellence Program ``Exploring New Science by Bridging
Particle-Matter Hierarchy'' of Tohoku University 
and partly by Monbukagakusho Scholarship 
and Grant-in-Aid for Scientific Research under
the program number 19740115 from the Japanese Ministry of
Education, Culture, Sports, Science and Technology.
\end{acknowledgments}

\appendix

\section{Numerical stabilization of coupled-channels calculation}

In this Appendix, we discuss the problem of numerical instability of
coupled-channels calculations and the stabilization methods which 
we employ in the present calculations. 

The coupled-channels equations (\ref{eq-multi-11}) 
form a set of $N$ second
order coupled linear differential equations, where $N$ is the
dimension of the coupled-channels equations. 
These equations can be solved 
by generating $N$ linearly independent solutions and taking a 
linear combination of these $N$ solutions so that the asymptotic 
boundary condition, (\ref{eq-scattwave}), as well as the regular 
boundary condition at the origin, 
are satisfied. 
The linearly independent solutions can be obtained by taking $N$ 
different sets of initial conditions at $r=0$. 
We denote these solutions by $\phi_{nn_i}(r)$, where $n$ refers to the 
channels while $n_i$ refers to a particular choice of the initial 
conditions. 
A simple choice for the $N$ initial conditions is to impose 
\begin{equation}
\phi_{nm}(r)\to cr^{J+1}\delta_{n,m},~~~~~{\rm for}~~r\to0, 
\end{equation}
where $c$ is an arbitrary number and $J$ is the total angular
momentum. With these boundary condition, the coupled-channels 
equations for $\phi_{nm}(r)$ given by 
\begin{eqnarray}
&&\bigg[-\frac{\hbar^2}{2\mu}\frac{d^2}{dr^2}+\frac{J(J+1)\hbar^2}{2\mu
  r^2}+V^{(0)}_N(r)+\frac{Z_PZ_Te^2}{r}\nonumber\\
&&-E+\epsilon_n\bigg]\phi_{nm}(r)+\sum_{n'}
  V_{nn'}(r)\phi_{n'm}(r)=0, 
\label{cc-phi} 
\end{eqnarray}
are solved outwards up to a matching radius $R_{\rm max}$. 
The wave functions $u_{n}(r)$ in Eq. (\ref{eq-multi-11}) are then 
obtained
as 
\begin{equation}
u_n(r)=\sum_m C_m\phi_{nm}(r),
\end{equation}
where the coefficients $C_m$ are determined so that the 
asymptotic boundary condition 
(\ref{eq-scattwave}) is fulfilled. 

In the classical forbidden region, the scattering 
wave functions exponentially
damp as the coordinate $r$ decreases. 
For the smaller energy, the damping is the stronger. Therefore, 
when the excitation energy $\epsilon_n$ is finite, 
the 
absolute value of the wave functions for each channel are different 
by order of magnitude in the classical forbidden region,  
and thus the wave functions 
tend to be dominated by that of the channel which has the smallest 
excitation energy. This easily destroys the linear independence of 
the $N$ numerical solutions $\phi_{nm}$, and causes the numerical
instability. This is a serious problem especially when the coupling 
is strong, as in the massive systems which we discuss in this paper. 

Several methods have been proposed in order to stabilize the
numerical solution of coupled-channels equations 
\cite{BBGR93,J78,L84,RO82}.
In the present calculations, we stabilize 
the solutions by diagonalizing the wave function matrix
$\phi_{nm}$ at several points of $r$ in order to recover the linear 
independence. 
That is, at some radius $r_s$, we compute the inverse of the matrix 
$A_{nm}=\phi_{nm}(r_s)$, and define the new set of wave functions,
\begin{equation}
\tilde{\phi}_{nm}(r)=\sum_k\phi_{nk}(r)\cdot (A^{-1})_{km}.
\label{stabilization}
\end{equation}
The new wave functions $\tilde{\phi}$ obey similar coupled-channels
equations as Eq. (\ref{cc-phi}), with the boundary conditions 
given by, 
\begin{eqnarray}
\tilde{\phi}_{nm}(r_s-h)&=&\sum_k\phi_{nk}(r-h)\cdot (A^{-1})_{km}, \\
\tilde{\phi}_{nm}(r_s)&=&\delta_{n,m}.
\end{eqnarray}
Here, $h$ is the step for the discretization of the radial coordinate,
$r$. 
These coupled-channels equations are solved outwards from $r_s$. 
The solutions $\phi$ can then be constructed as
$\phi=A\cdot \tilde{\phi}$. 
We impose this stabilization procedure up to $r=15$ fm with an
interval of 1 fm. 
Although this method is similar to those in Refs. \cite{L84,RO82}, our 
method is much simpler to be implemented. 

This method is sufficient for intermediate heavy systems, such as
$^{16}$O + $^{144}$Sm. For massive systems, however, we still 
encounter a small numerical instability \cite{H06}. 
In order to cure this problem, in addition to the stabilization method 
(\ref{stabilization}), we also adopt two other methods, which are used in
the computer code {\tt FRESCO}\cite{T88}. 
That is, we introduce two radii, $R_{\rm min}$ and $R_{\rm cut}$. 
$R_{\rm min}$ is the radius from which 
the coupled-channels equations (\ref{cc-phi}) are solved, {\it i.e.},
these equations are solved from $r=R_{\rm min}$ instead of $r=0$, 
by setting $\phi_{nm}(r)=0$ for $r\leq R_{\rm min}$. 
$R_{\rm cut}$ is a cut-off radius for the coupling matrix, {\it i.e.}, 
the off-diagonal components of the coupling matrix $V_{nn'}(r)$ are 
set to be zero for $r\leq R_{\rm cut}$. 
Both the procedures are justified when the absorption is strong inside
the Coulomb barrier, as in heavy-ion systems, 
and the results are insensitive to the particular choice of 
$R_{\rm min}$ and $R_{\rm cut}$ as long as they are inside the Coulomb 
barrier. Typically, we take 
$R_{\rm min}$= 6 fm  and $R_{\rm cut}$= 10 fm to obtain reasonable
results for the present systems (notice that the pocket and the
barrier appear at {\it e.g.,} 11.3 and 13.2 fm, respectively, for 
the $^{64}$Ni+$^{208}$Pb system with the nuclear potential given in
Table I).


\begin{thebibliography}{109}

\bibitem{bah-tak-98} A.B. Balantekin and N. Takigawa, Rev. Mod. Phys.
 \textbf{70}, 77 (1998). 

\bibitem{das-98} M. Dasgupta, D.J. Hinde, N. Rowley and A.M. Stefanini, 
Annu. Rev. Part. Sci. \textbf{48}, 401 (1998).

\bibitem{daso-83} C.H. Dasso, S. Landowne and A. Winther,
  Nucl. Phys. \textbf{A405}, 381 (1983); \textbf{A407}, 221 (1983). 

\bibitem{HTB97}K. Hagino, N. Takigawa, and A.B. Balantekin, 
Phys. Rev. {\bf C56}, 2104 (1997). 

\bibitem{hag-bah-04} K. Hagino and A.B. Balantekin,
  Phys. Rev. \textbf{A70}, 032106 (2004). 

\bibitem{row-91} N. Rowley, G.R. Sacthler and P.H. Stelson,
  Phys. Lett. \textbf{B254}, 25 (1991).

\bibitem{leigh-95} J.R. Leigh, M. Dasgupta, D.J. Hinde, 
J.C. Mein, C.R. Morton, J.P. Lestone, J.O. Newton, H. Timmers,
J.X. Wei and N. Rowley, Phys. Rev. \textbf{C52}, 3151 (1995).

\bibitem{row-06} N. Rowley, N. Grar and K.Hagino, Phys. Lett. 
\textbf{B632}, 243 (2006).

\bibitem{thim-95} H. Timmers, J.R. Leigh, M. Dasgupta, D.J. Hinde, 
R.C. Lemon, J.C. Mein, C.R. Morton, J.O. Newton and N. Rowley, Nucl. Phys. 
\textbf{A584}, 190 (1995).

\bibitem{hag-04} K.Hagino and N. Rowley, Phys. Rev. \textbf{C69}, 
054610 (2004).

\bibitem{andres-88} M.V. Andres, N. Rowley and M.A. Nagarajan, 
Phys. Lett. \textbf{B202}, 292 (1988).

\bibitem{zam-07}Muhammad Zamrun F. and K. Hagino, arXiv:0710.2753. 

\bibitem{row-061} N. Rowley {\it et al.}, Phys. At. Nucl.,
  \textbf{79}, 1093 (2006) 

\bibitem{ntsha-07} S.S. Ntshangase {\it et al.},
  Phys. Lett. \textbf{B651}, 27 (2007). 

\bibitem{ike-06} H. Ikezoe {\it et al.}, AIP Conf. Proc.,
  \textbf{853}, 69 (2006). 

\bibitem{mitsu-07} S. Mitsuoka, H. Ikezoe, K. Nishio, K. Tsuruta, S.C. Jeong 
and Y. Watanabe, Phys. Rev. Lett. {\bf 99}, 182701 (2007).


\bibitem{hag-99} K. Hagino, N. Rowley and A.T. Kruppa,
  Comput. Phys. Commun. \textbf{123}, 143 (1999).

\bibitem{es-lan-pri-87} H. Esbensen, S. Landowne and H. Price,
  Phys. Rev. {\bf C36}, 1216 (1987); {\bf 36}, 2359 (1987).

\bibitem{hag2} K. Hagino {\it et al.} (to be published).

\bibitem{washi-06} K. Washiyama, K. Hagino and M. Dasgupta,
  Phys. Rev. \textbf{C73}, 034607 (2006). 

\bibitem{gasques-07} L.R. Gasques, M. Evers, D.J. Hinde, M. Dasgupta,
  P.R.S. Gomes, R.M. Anjos, M.L. Brown, M.D. Rodriguez, R.G. Thomas
  and K. Hagino,  Phys. Rev. \textbf{C76}, 024612 (2007).

\bibitem{hinde-07}
D.J. Hinde, R.L. Ahlefeldt, R.G. Thomas, K. Hagino, M.L. Brown, 
M. Dasgupta, M. Evers, L.R. Gasques, and M.D. Rodriguez,
Phys. Rev. C{\bf 76}, 014617 (2007). 

\bibitem{spear-02} T. Kibedi and R.H. Spears, At. Data and Nucl. Data
  Tables, \textbf{80}, 35 (2002).

\bibitem{raman-01} S. Raman, C.W. Nestor and P. Tikkanen, At. Data and
  Nucl. Data Tables, \textbf{78}, 1 (2001).

\bibitem{das-97} M. Dasgupta, K. Hagino, C.R. Morton, D.J. Hinde,
  J.R. Leigh, N. Takigawa, H. Timmers, and J.O. Newton,
  Jour. Phys. {\bf G23}, 1491 (1997).

\bibitem{morton-99} C.R. Morton, A.C. Berriman, D.J. Hinde,
  J.O. Newton, K. Hagino and I.J. Thompson, Phys. Rev. \textbf{C60},
  044608 (1999).

\bibitem{LINDA}E.Duek, L. Kowalsky and John M. Alexander, Comp. Phys.
Comm. {\bf 34}, 395 (1985)

\bibitem{hag-97} K. Hagino, N. Takigawa and S. Kuyucak, Phys. Rev. Lett.
\textbf{79}, 2943 (1997).

\bibitem{hag-971} K. Hagino, S. Kuyucak and N. Takigawa, Phys. Rev.
\textbf{C57}, 1349 (1997).




\bibitem{brogwin-91}R.A. Broglia and A. Winther, {\it Heavy Ion Reactions}, 
Vol. {\bf 84} in Frontier in Physics Lecture Notes Series 
(Addison-Wesley, Redwood City, CA, 1991).

\bibitem{criswin-76}P.R. Christensen and A. Winther, Phys. Lett. 
{\bf B65}, 19 (1976).

\bibitem{satc-79}G.R. Sacthler and W.G. Lowe, Phys. Rep. {\bf 55}, 183
  (1979).

\bibitem{NBD04} J.O. Newton, R.D. Butt, M. Dasgupta, D.J. Hinde,
  I.I. Gontchar, C.R. Morton and K. Hagino, Phys. Rev. {\bf C70},
  024605 (2004); Phys. Lett. {\bf B586}, 219 (2004).







\bibitem{BBGR93}W. Brenig, T. Brunner, A. Gross, and R. Russ,
Z. Phys. {\bf B93}, 91 (1993); W. Brenig and R. Russ,
Surf. Sci. {\bf 315}, 195 (1994).

\bibitem{J78}B.R. Johnson, J. Chem. Phys. {\bf 69}, 4678 (1978).

\bibitem{L84}Z.H. Levine, Phys. Rev. A{\bf 30}, 1120 (1984).

\bibitem{RO82}T.N. Rescigno and A.E. Orel, Phys. Rev. A{\bf 25},
2402 (1982).

\bibitem{H06}K. Hagino, AIP Conf. Proc. {\bf 891}, 80 (2006); 
arXiv: nucl-th/0611015. 

\bibitem{T88}I.J. Thompson, Comp. Phys. Rep. {\bf 7}, 167 (1988).

\end{thebibliography}
\end{document}